\documentclass[aps,prb,reprint,superscriptaddress,floatfix]{revtex4-2}
\usepackage[normalem]{ulem}
\usepackage{natbib}
\usepackage{xcolor}
\usepackage{graphicx}
\usepackage{amsmath}
\usepackage{physics}
\usepackage[colorlinks=true,allcolors=blue, pdftitle={BFCA}]{hyperref}%
\DeclareUnicodeCharacter{2212}{-}

\usepackage{xspace}

\newcommand{\bfca}{Ba(Fe$_{1-x}$Co$_{x}$)$_2$As$_2$\xspace}

\newcommand{\Aog}{A$_\mathrm{1g}$\xspace}
\newcommand{\Bog}{B$_\mathrm{1g}$\xspace}
\newcommand{\Tc}{$T_\mathrm{c}$\xspace}
\newcommand{\Tce}{T_\mathrm{c}}
\newcommand{\TMFL}{$T^\ast$\xspace}

\newcommand{\etal}{\textit{et al.}\xspace}

\newcommand{\xc}{$x_\mathrm{c}$\xspace}
\newcommand{\xce}{x_\mathrm{c}\xspace}
\newcommand{\Tnem}{$T_\mathrm{nem}$\xspace}
\newcommand{\Tneme}{T_\mathrm{nem}}
\newcommand{\BLT}{$\beta\tilde\Lambda(\beta/2)$\xspace}

\renewcommand{\sout}[1]{\unskip}

\begin{document}

\title{Quantum critical fluctuations in an Fe-based superconductor}
\date{\today}
\author{Daniel Jost}
\email{daniel.jost@stanford.edu}
\affiliation{Walther Meissner Institut, Bayerische Akademie der Wissenschaften, 85748 Garching, Germany}
\affiliation{Fakult\"at f\"ur Physik, Technische Universit\"at M\"unchen, 85748 Garching, Germany}
\affiliation{Stanford Institute for Materials and Energy Sciences,
SLAC National Accelerator Laboratory, 2575 Sand Hill Road, Menlo Park, CA 94025, USA}
\author{Leander Peis}
\altaffiliation{Current address: IFW Dresden, Helmholtzstr. 20, 01069 Dresden, Germany}
\affiliation{Walther Meissner Institut, Bayerische Akademie der Wissenschaften, 85748 Garching, Germany}
\affiliation{Fakult\"at f\"ur Physik, Technische Universit\"at M\"unchen, 85748 Garching, Germany}
\author{Ge He}
\affiliation{Walther Meissner Institut, Bayerische Akademie der Wissenschaften, 85748 Garching, Germany}
\author{Andreas Baum}
\altaffiliation{Current address: Mynaric, Dornier Str. 19, 82205 Gilching, Germany}
\affiliation{Walther Meissner Institut, Bayerische Akademie der Wissenschaften, 85748 Garching, Germany}
\author{Stephan Geprägs}
\affiliation{Walther Meissner Institut, Bayerische Akademie der Wissenschaften, 85748 Garching, Germany}
\author{Johanna\,C.~Palmstrom}
\altaffiliation{Current address: Los Alamos National Laboratory, Los Alamos, NM 87545, USA}
\affiliation{Stanford Institute for Materials and Energy Sciences,
SLAC National Accelerator Laboratory, 2575 Sand Hill Road, Menlo Park, CA 94025, USA}
\affiliation{Geballe Laboratory for Advanced Materials \& Dept. of Applied Physics,
Stanford University, CA 94305, USA}
\author{Matthias\,S.\,Ikeda}
\affiliation{Stanford Institute for Materials and Energy Sciences,
SLAC National Accelerator Laboratory, 2575 Sand Hill Road, Menlo Park, CA 94025, USA}
\affiliation{Geballe Laboratory for Advanced Materials \& Dept. of Applied Physics,
Stanford University, CA 94305, USA}
\author{Ian\,R. Fisher}
\affiliation{Stanford Institute for Materials and Energy Sciences,
SLAC National Accelerator Laboratory, 2575 Sand Hill Road, Menlo Park, CA 94025, USA}
\affiliation{Geballe Laboratory for Advanced Materials \& Dept. of Applied Physics,
Stanford University, CA 94305, USA}
\author{Thomas Wolf}
\affiliation{Karlsruhe Institut für Technologie, P.O. Box 3540, 76021 Karlsruhe, Germany}
\author{Samuel~Lederer}
\affiliation{Cornell University, Ithaca, New York 14850, USA}
\author{Steven\,A.~Kivelson}
\affiliation{Department of Physics, Stanford University, Stanford, CA 94305, USA}
\author{Rudi~Hackl}
\email{hackl@tum.de}
\affiliation{Walther Meissner Institut, Bayerische Akademie der Wissenschaften, 85748 Garching, Germany}
\affiliation{Fakult\"at f\"ur Physik, Technische Universit\"at M\"unchen, 85748 Garching, Germany}
\affiliation{IFW Dresden, Helmholtzstr. 20, 01069 Dresden, Germany}

\date{\today}


\begin{abstract}
Quantum critical fluctuations may prove to play an instrumental role in the formation of unconventional superconductivity. Here, we show that the characteristic scaling of a marginal Fermi liquid is present in inelastic light scattering data of an Fe-based superconductor tuned through a quantum critical point (QCP) by chemical substitution or doping. From the doping dependence of the imaginary time dynamics we are able to distinguish regions dominated by quantum critical behavior from those having classical critical responses. This dichotomy reveals a connection between the marginal Fermi liquid behavior and quantum criticality. In particular, the overlap between regions of high superconducting transition temperatures and quantum critical scaling suggests a contribution from quantum fluctuations to the formation of superconductivity.
\end{abstract}

\maketitle

\section{Introduction}

Close to quantum critical points (QCPs) where a second order thermal phase transition is suppressed to absolute zero systems are highly susceptible to small perturbations. The high susceptibility gives rise to various instabilities in materials exhibiting unconventional superconductivity. Although the vanishing temperature at the QCP renders direct experiments impossible, the related critical fluctuations \cite{Hertz:1976,Coleman:2005} extend over a wide range of temperatures and manifest themselves under experimentally feasible conditions, yielding direct consequences for the macroscopic properties of materials.

One of the consequences is the deviation from the Landau-Fermi liquid picture observed in many materials such as the cuprates, the Fe-based compounds or the heavy fermion systems \cite{Fradkin:2015}. The linear-in $T$ variation of the resistivity has been considered a landmark experimental manifestation \cite{Fournier:1998,Jin:2011,Martin:1990,Cooper:2009,Licciardello:2019} thereof \cite{Varma:1989}, although the interrelation of the resistivity with quantum criticality remains a subject of discussion \cite{Murthy:2021}.
On the other hand, spectroscopic fingerprints strongly indicate a connection between the $\Omega/T$ scaling of a marginal Fermi liquid \cite{Varma:1989} and quantum criticality, as observed in a recent study of the optical conductivity \cite{Prochaska:2020} of the heavy fermion superconductor YbRh$_2$Si$_2$ \cite{Schuberth:2016}. Theoretical works suggest that quantum fluctuations may have a significant impact in the case for antiferromagnetic \cite{Scalapino:2012} as well as charge density wave ordering \cite{Perali:1996} and, more specifically, in the enhancement \cite{Lederer:2015} or formation \cite{Wang:2016, Lederer:2020:PRR} of superconductivity in the iron-based superconductors.

Characteristic for systems with a QCP is the presence of a quantum critical region at temperatures above a cross-over temperature $T^{\ast}(x) \sim |x-\xce|^{\nu z}$ (where  $\nu$ and $z$ are the correlation length and dynamical critical exponents, respectively, the tuning parameter $x$, realized in our context by chemical substitution of the metal site, and critical doping \xc). Since this broad region of the phase diagram manifests the behavior of the single quantum critical point at $x=x_c,\ T=0$, Frerot \etal \cite{Frerot:2019} dubbed this region a ``magnifying lens'' for quantum criticality. 

Due to the tunability by doping and the purity of the crystals, we have opted to use as a testbed the compound \bfca, one of the most studied Fe-based superconductors, to investigate this lens. \bfca  exhibits a rich phase diagram \cite{Chu:2009} with overlapping regions of magnetism, nematicity and superconductivity, and shows a strong enhancement of nematic fluctuations having $d_{x^2-y^2}$ symmetry (here using the 1-Fe unit cell) close to the nematic transition \cite{Gallais:2013, Gallais:2016, Kretzschmar:2016}. A plethora of experimental studies \cite{Gallais:2013,Gallais:2016,Chu:2012,Boehmer:2014,Kuo:2016,Palmstrom:2022} indicate the existence of a QCP underneath the superconducting dome in this material, and the Raman response within the superconducting phase bears an imprint of nematic quantum criticality \cite{Gallais:2016_PRL}. Most recently, Worasaran \etal \cite{Worasaran:2021} provided striking evidence for nematic quantum criticality by observing power-law behavior of the nematic/structural transition temperature as a function of uniaxial strain. However, the influence of the QCP on the high temperature characteristics, for instance signatures of $\Omega/T$ scaling, remain elusive. \newline
In this Article we provide a comprehensive investigation of the Raman response of \bfca for doping concentrations on both sides of optimal doping $\xce \sim 0.061$. Specifically, we fill the gap close to \xc where $\Omega/T$ scaling is expected to be most pronounced and provide a solid experimental basis for an earlier proposal \cite{Lederer:2020}. We use electronic inelastic light (Raman) scattering which probes fluctuations of various degrees of freedom \cite{Aslamasov:1968,Caprara:2005} and has led to promising results particularly in Fe-based superconductors \cite{Gallais:2013,Thorsmolle:2016, Kretzschmar:2016}. 

\section{Results and discussion}
Earlier studies, investigating the low frequency Raman response of \bfca \cite{Gallais:2013,Kretzschmar:2016}, showed that the contribution of nematic fluctuations presents itself in the \Bog ($d_{x^2-y^2}$) symmetry of the 1-Fe unit cell whereas the temperature dependence of the \Aog response is essentially independent of doping and follows the in-plane resistivity \cite{Kretzschmar:2016}. This suggests, that the essential critical dynamics of the system is captured by the \Bog response. We therefore  limit our analysis to the \Bog response, with characteristic spectra shown in Fig. \ref{fig:5.5} for selected \bfca samples with $x<\xce$, $x\sim \xce$ and $x\gg\xce$ (see Fig. S1 and S2 in Supplementary Note 1 for the complete data set). All spectra have in common that the temperature dependence is limited to low frequencies, while no significant temperature dependence can be detected  at higher frequencies.
\begin{figure}
  \centering
  \includegraphics[width=85mm]{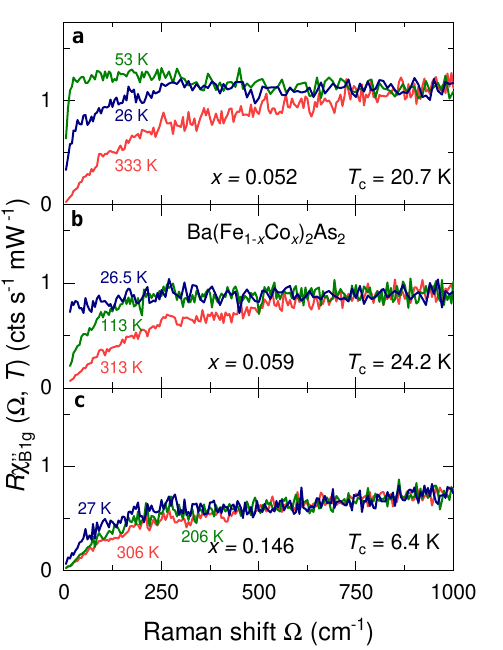}
  \caption{Temperature and doping dependence of the Raman response. Depicted is the Raman response function $R\chi_\mathrm{B1g}^{\prime\prime}(\Omega,T)$ in \Bog symmetry (1-Fe unit cell) at selected doping concentrations $x$ and temperatures $T$ where $\Omega = \omega_\mathrm{i} - \omega_\mathrm{s}$ stands for the difference of the incident and scattered photon energy $\omega_\mathrm{i}$ and $\omega_\mathrm{s}$, respectively. \Tc refers to the superconducting transition temperature. \textbf{a} For the underdoped compound with $x = 0.052$, the low-frequency pile-up attributable to classical critical fluctuations exhibits a maximum above the nematic transition and decreases towards \Tc. \textbf{b} Close to optimal doping with $x=0.059$ the low frequency response increases monotonically on cooling towards \Tc. \textbf{c} For a sample with $x\gg x_c$, The temperature dependence is weak at all frequencies.}
  \label{fig:5.5}
\end{figure}

\subsection{Raman response}
The underdoped compound [Fig. \ref{fig:5.5} \textbf{a}] exhibits a nematic transition at approximately $\Tneme \sim 50\,\mathrm{K}$. Similar to results reported elsewhere for $x<\xce$ \cite{Gallais:2013,Thorsmolle:2016,Kretzschmar:2016} the strong increase of the spectral weight between $333$ and $53\,\mathrm{K}$ in the range below $500\,\mathrm{cm}^{-1}$ can be attributed to the presence of (classical) critical fluctuations. The decrease below \Tnem, reminiscent of a gap, is characteristic for such transitions in the Fe-based superconductors, and consistent with earlier results \cite{Kretzschmar:2016}. Closer to optimal doping at $x=0.059$ [Fig. \ref{fig:5.5} \textbf{b}] the response increases monotonically upon cooling towards the superconducting transition temperature \Tc. Remarkably, the response at this lowest temperature is nearly constant over the entire frequency range. Far away from \xc, the temperature dependence is weak even at small energy transfer [Fig. \ref{fig:5.5} \textbf{c}].

\subsection{Transformation onto the imaginary time axis}
\begin{figure}
  \centering
  \includegraphics[width=85mm]{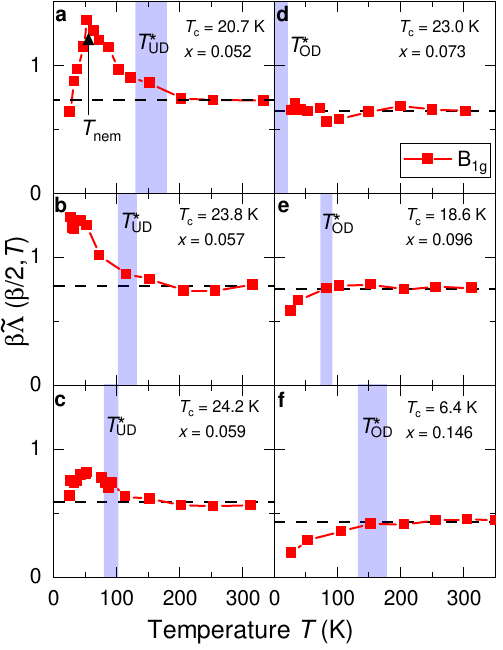}
  \caption{Transformation onto the imaginary time axis. Shown is $\beta\tilde{\Lambda}(\beta/2)$, the imaginary time-ordered correlation function  
  $\tilde{\Lambda}(\tau)$ derived from the Raman spectra of \bfca and evaluated at $\tau = \beta/2$ multiplied by the inverse temperature $\beta = 1/k_\mathrm{B}T$, versus temperature $T$. The dashed lines correspond to $\beta\tilde{\Lambda}(\beta/2,T)= \mathrm{const}$, and the blue regions indicate the deviation from this constant behavior, which we use to define the crossover temperatures $T^\ast_\mathrm{UD}$ and $T^\ast_\mathrm{OD}$ for $x<\xce$ and $x>\xce$, respectively. \textbf{a} At $x = 0.052$ doping a nematic transition occurs at \Tnem $\sim$ 50\,K at which $\beta\tilde{\Lambda}(\beta/2,T)$ develops a cusp and then decreases towards \Tc. \textbf{b} At $x = 0.057$, $\beta\tilde{\Lambda}(\beta/2,T)$ increases towards lower temperatures and develops a maximum above $T_\mathrm{c}$. \textbf{c} Closer to the quantum critical point (QCP) $\beta\tilde{\Lambda}(\beta/2,T)$ develops a hump between $T^\ast_\mathrm{UD}$ and \Tc. \textbf{d} On the overdoped side at $x = 0.073$ the $\Omega/T$ scaling persists down to \Tc. \textbf{e} Further away, the crossover temperature increases again. \textbf{f} The high crossover temperature at $x=14.6$ coincides with a small superconducting transition temperature.}
  \label{fig:bL}
\end{figure}

To examine the possibility of quantum critical scaling, we have computed \BLT from each of our measured spectra, with the results shown in Fig. \ref{fig:bL}. For the underdoped sample [Fig. \ref{fig:bL} \textbf{a}] the characteristic cusp singularity of the nematic transition is visible \cite{Lederer:2020}. Notably, a similar increase observed using a sample with $x=0.057$ doping in Fig. \ref{fig:bL} \textbf{b} towards $T_\mathrm{c}$ suggests a persisting nematic phase close to the superconducting transition. Yet, from our measurements it cannot be clearly decided whether or not nematicity sets in above \Tc. Closer to optimal doping in Fig.~\ref{fig:bL}~\textbf{c}, $\beta\tilde{\Lambda}(\beta/2)$ develops a broad maximum above \Tc, though this cannot be attributed to a nematic phase transition. This maximum occurs roughly at the same temperature as for 5.2\% doping and is thus too high in temperature for this doping concentration. Rather, we believe that the increase in the critical fluctuation response starts to freeze out at low temperatures and is then proceedingly compensated by the denominator in eq. \ref{eqn:BLT}. Above $x_\mathrm{c}$ at $x=0.073$ [Fig.~\ref{fig:bL}~\textbf{d}], the quantity $\beta\tilde{\Lambda}(\beta/2)$ remains constant down to the lowest measured temperature on the order of \Tc. Tuning further away from \xc [Fig. \ref{fig:bL} \textbf{e} and \textbf{f}], $\beta\tilde{\Lambda}(\beta/2)$ decreases towards low temperatures, where \BLT is generically required to vanish (see the Supplementary Note 2 for a proof of this).

\subsection{Doping dependence of \TMFL}

The dashed horizontal line is a guide to the eye indicating the value of $\beta\tilde{\Lambda}$ in the regime of $\Omega/T$ scaling. Outside this quantum critical regime, where the response does not scale as $\Omega/T$, the quantity $\beta\tilde{\Lambda}(\beta/2)$ from this constant. The shaded regions show the crossover temperatures $T^\ast_\mathrm{UD}$ and $T^\ast_\mathrm{OD}$ for $x<\xce$ and $x>\xce$, respectively, from $\Omega/T$ scaling to regions where other dynamics govern the low frequency response.

\begin{figure}
  \centering
  \includegraphics[width=85mm]{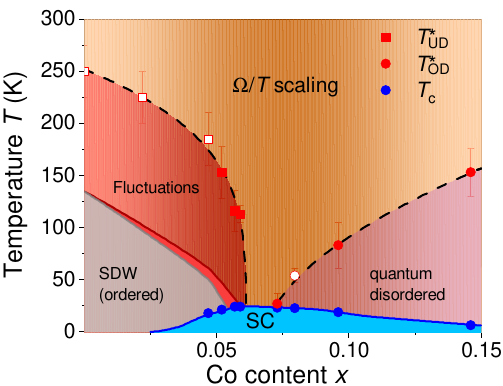}
  \caption{Phase diagram of \bfca. The spin-density-wave (SDW) phase is depicted in grey, nematicity in red and superconductivity (SC) in blue as defined by the superconducting transition temperature \Tc. The rectangular and circular data points indicate the cross-over temperatures $T^\ast_\mathrm{UD}$ and $T^\ast_\mathrm{OD}$ from the region of $\Omega/T$ scaling for $x<\xce$ and $x>\xce$, respectively. The dashed lines are guides to the eye. Open red squares and circles represent data points from Lederer \etal \cite{Lederer:2020}. The grey, red and blue phase separation lines are based on Chu \etal \cite{Chu:2009}. The error bars indicate the temperature window of $T^\ast$ depicted in Fig.~\ref{fig:bL}.}
  \label{fig:PD}
\end{figure}

$T^\ast_\mathrm{UD,OD}(x)$ exhibits a systematic doping dependence, with a global minimum at $x=0.073$. To further illustrate this point, $T^\ast_\mathrm{UD,OD}(x)$ is plotted along with the relevant phase transitions of \bfca in Fig. \ref{fig:PD}. The spin-density-wave (SDW) phase is indicated in grey, nematicity in red, and superconductivity in blue. The open red rectangles indicate the crossover temperatures derived from data presented in Lederer \etal \cite{Lederer:2020}. The dashed lines represent guides to the eye.

\textbf{Underdoped region:}  For $x<\xce$ \bfca has an SDW ground state below $T_\mathrm{SDW}$, and fluctuations in the temperature range $T_\mathrm{SDW}<T<T^\ast_\mathrm{UD}$. Above $T^\ast_\mathrm{UD}$ the $\Omega/T$ scaling characteristic for a marginal Fermi liquid is recovered. Moving closer to \xc, the magnetic and nematic phases become unstable as signified by the decreasing transition temperatures. In the picture of quantum criticality, this results from a continously increasing superposition of entangled states which manifests itself as a divergence of the coherence length of the quantum fluctuations directly at the QCP. Simultaneously, the region of marginal Fermi liquid behavior with $\Omega/T$ scaling begins to persist over wider ranges of temperature with $T^\ast_\mathrm{UD}$ falling more rapidly after superconductivity sets in.

\textbf{Overdoped region:} On the overdoped side for $x>\xce$, the crossover line re-emerges from the superconducting dome and continues to rise as a function of doping, with a significantly smaller slope than for $x<\xce$. In the absence of a competing magnetic ground state, the quantum fluctuations may support superconductivity over a wider region of doping concentrations.

The endpoints of the dashed lines coincide with the region of optimal doping \xc consistent with a scenario in which quantum fluctuations increasingly destroy the long range order of the ground state for $x<\xce$ at temperatures $T>0$. Thus, \Tc is high where the quantum fluctuations dominate the normal state, and low where "classical" dynamics take over. Therefore the quantum fluctuations appear to have a twofold effect: on the classically ordered side below \xc, the competing magnetic ground state is destabilised. On the overdoped side, the quantum fluctuations  protect superconductivity at least partially from increasing disorder. This scenario is consistent with quantum criticality being a driving force behind the superconducting state in \bfca.

In the experimental work presented here, we have determined the crossover temperature $T^\ast$ separating the high temperature quantum critical behavior from the (effectively) classical behavior at low temperatures. 
To this end, we have introduced a metric to identify the quantum critical regime in terms of the imaginary time response function,  $\tilde{\Lambda}(\tau)$, which can be inferred directly and unambiguously from experimental inelastic light scattering (Raman) data (see Methods \ref{sec:imtimeresp} and Lederer \etal \cite{Lederer:2020}). By focusing on the longest possible imaginary time, $\tau = \beta/2$ with the inverse temperature $\beta = 1/k_\mathrm{B}T$, we amplify the effect of the singular (power-law) correlations that reflect the critical behavior and maximally avoid contamination by non-universal analytic (short-range in imaginary time) non-critical portions of the imaginary part of the response function $\chi^{\prime\prime}$. Specifically, if in the critical regime $\chi^{\prime\prime}$ is a scaling function of $\Omega/T$ (i.e. of the marginal Fermi liquid form), this implies that $\tilde{\Lambda}(\beta/2)$ is linear in $T$.  In this way, as shown in Fig. \ref{fig:bL}, we can identify the quantum critical regime for each value of $x$ as the range of (high) $T$ over which $\tilde{\Lambda}(\beta/2)$ is $T$ independent, and a crossover scale $T^\ast(x)$ that bounds this regime from below.  Applying this analysis to the present work leads to the phase diagram shown in Fig. \ref{fig:PD}, where for $x<\xce$, the crossover is {in}to a renormalized classical regime in which $\beta\tilde{\Lambda}(\beta/2)$ is an increasing function of decreasing $T$, while for $x >\xce$ the crossover is {in}to a quantum disordered regime in which  $\beta\tilde{\Lambda}(\beta/2)$ is a decreasing function.

\subsection{Scaling}
 As a final remark, we wish to comment on the scaling behavior deduced from our data: From equation (S12) in Supplementary Note 2.4, $\beta\tilde{\Lambda}(\beta/2)$ should vary as $T^y$ where $y = \gamma/\nu z$. This implies $y \sim 0$ to be in line with a conventional scaling analysis presented in Supplementary Note 2, and thus one would have to appeal to an extreme value of one or more of these exponents. In this case the most natural choice would be $z=\infty$, corresponding to the phenomenology of local quantum criticality \cite{Si:2001, Varma:2020} in which only temporal correlations become long ranged at the quantum critical point. Such a scenario is of course exotic. Whether a local quantum critical point that violates scaling is indeed present in \bfca is beyond the scope of this work and calls for further investigations of quantum criticality in this and related systems.

\section{Conclusions}
We tracked the \Bog Raman response of \bfca as a function of doping $x$. $x$ is tuned from $x<\xce$ through a quantum critical point far into the overdoped regime of the phase diagram. From the data we calculated the temperature dependence of the imaginary time ordered correlation function $\tilde\Lambda(\tau)$ at $\tau=\beta/2$ and extracted cross-over temperatures $T^\ast(x)$ above which $\Omega/T$ scaling prevails. These temperatures define the boundaries of the quantum critical fan corroborating the existence of a QCP around optimal doping. Thereby we show that the determination of the imaginary time dynamics from spectroscopic data is a straightforward way to reveal the effects of quantum criticality in the normal state of \bfca. The most important distinction is drawn between this analysis and any type of Kramers-Kronig analysis where an upper limit needs to be set. Here, the integral is naturally cut off  at $k_\mathrm{B} T$ by an exponential factor. The analysis is not limited to inelastic light scattering, but can be applied to a wide range of experimental probes as already outlined in Lederer \etal \cite{Lederer:2020}. We hope that this transformation may enable the comparison of results from different spectroscopic probes directly and on the same footing, allowing a more unified analysis of quantum criticality.

\section{Methods}
\subsection{Raman scattering}
We performed inelastic light scattering experiments with the samples (for characterization, see Supplementary Note 3) attached to the cold finger of a $^4$He flow cryostat. Polarized photons having a wavelength of 577~nm (Coherent GENESIS MX-SLM577-500) hit the sample at an angle of incidence of $68^\circ$. The polarized scattered photons were collected along the surface normal of the sample and focused on the entrance slit of a double monochromator. The photons transmitted at the selected energy were recorded with a CCD detector. The number of photons per second is proportional to the Van-Hove function $S(\mathbf{q}\approx 0, \Omega) = 1/\pi [{1+n(\Omega,T)}]\chi^{\prime\prime}(\Omega,T)$ where $n(\Omega,T)$ is the Bose factor and $\chi^{\prime\prime}$ is the imaginary part of Raman response function which is displayed in Fig.~\ref{fig:5.5}. The factor $R$ is the constant of proportionality which absorbs all experimental factors.

\subsection{Imaginary time response}
\label{sec:imtimeresp}
Following a method proposed in Lederer \etal \cite{Lederer:2020}, we have used our measurements of the dissipative part of the response function, $\chi^{\prime\prime}$, to characterize the dynamics of the corresponding \emph{imaginary time}  correlation function $\tilde{\Lambda}(\tau)$ using the following exact identity:
\begin{equation}
\tilde{\Lambda}(\tau)= \int\frac{\mathrm{d}\Omega}{2\pi} \chi^{\prime\prime}(\Omega,T) \frac{\exp[\Omega(\tau-\beta/2)]}{\sinh[\beta\Omega/2]},
\label{eqn:lambda}
\end{equation}
with the imaginary time $\tau \in (0,\beta)$ and the inverse temperature $\beta =1/T$ where $k_\mathrm{B} = 1$ along the lines of Lederer \etal \cite{Lederer:2020}. Since bosonic correlations have period $\beta$ in imaginary time, the longest time response corresponds to $\tau\approx\beta/2$. In the following, it will be convenient to work with the specific quantity
\begin{equation}
\beta\tilde{\Lambda}(\beta/2)= \int\frac{\mathrm{d}\Omega}{2\pi T} \frac{\chi^{\prime\prime}(\Omega,T) }{\sinh[\beta\Omega/2]},
\label{eqn:BLT}
\end{equation}
which has the same universal properties as the static susceptibility $\chi^{\prime}(\Omega=0,T)$ in the sense that the two quantities have identical scaling with temperature on cooling to both classical and quantum critical points \cite{Lederer:2020}.

A key property of \BLT is that it has a power-law dependence on temperature if $\chi^{\prime \prime}$ obeys $\Omega/T$ scaling, and in particular is \emph{independent of temperature} if $\chi^{\prime\prime}(\Omega,T) $ is of marginal Fermi liquid form \cite{Varma:1989}, i.e. if it is a function of only $\Omega/T$ and not of either variable separately: 
\begin{equation}
\mathrm{Im}\bar{P}(\mathbf{q}\rightarrow 0, \Omega) \sim \chi^{\prime\prime}(\Omega) \sim \frac{\Omega}{\mathrm{max}(\Omega,T)}.
\label{eqn:scale}
\end{equation}

One significant virtue of \BLT in comparison to $\chi\prime(\Omega=0)$ is that it can be unambiguously computed from the measured Raman spectra. Since the measured $\chi''$ (in this and other studies) fails to fall off at high frequency, the Kramers-Kronig integral relating $\chi''$ to $\chi'$ does not converge, and therefore requires a manual cutoff procedure. By contrast, the denominator in Eq. \ref{eqn:BLT} grows exponentially for frequencies $\Omega\gg T$, rendering the integral convergent so long as the range of frequencies probed extends to several times the temperature.

\newpage

\section*{Data availability statement} 
Data are available from the corresponding authors upon reasonable request.

\section*{Author contributions} 
D.J., L.P., G.H. and A.B. contributed approximately equally to the experiments. D.J. and L.P. performed the analysis. J.C.P., M.I., I.R.F. and T.W. prepared and characterized the samples. S.G. performed the SQUID measurements. S.L. worked out the theory. D.J. and R.H. conceived the study and prepared the manuscript with input from all authors, in particular from S.A.K..

\section*{Competing financial interests} 
The authors declare no competing interests.

\begin{acknowledgments}
We thank E. Berg and B. Moritz for fruitful discussions. The work was supported by the Deutsche Forschungsgemeinschaft (DFG, Projekt-IDs 107745057 – TRR 80,  HA 2071/8-1 and HA2071/12-1)), the Bavaria-California Technology Center (BaCaTeC, grant no. 21[2016-2]), the Friedrich-Ebert foundation (D.J.) and the Alexander-von-Humboldt foundation (G.H., D.J.). Part of the analysis was performed at the Stanford Institute for Materials and Energy Sciences (SIMES) funded by the U.S. Department of Energy, Office of Basic Energy Sciences, Materials Sciences and Engineering Division (D.J.). J. C. P. was supported by a Gabilan Stanford Graduate Fellowship and a Stanford Lieberman Fellowship. S. L. is supported by the U.S. Department of Energy, Office of Science, National Quantum Information Science Research Centers, Quantum Systems Accelerator (QSA).
\end{acknowledgments}

\clearpage
\newpage

\renewcommand{\thefigure}{S\arabic{figure}} 
\renewcommand{\theequation}{S\arabic{equation}}
\renewcommand{\thesection}{SUPPLEMENTARY NOTE \arabic{section}}  
\renewcommand{\thesubsection}{SUPPLEMENTARY NOTE \arabic{section}.\arabic{subsection}}  
\renewcommand{\thetable}{ST\arabic{table}}   
\renewcommand{\tablename}{Supplementary table} 

\setcounter{figure}{0}  
\setcounter{section}{0}

\section{Raman data}

\begin{figure}[ht]
  \centering
  \includegraphics[width=85mm]{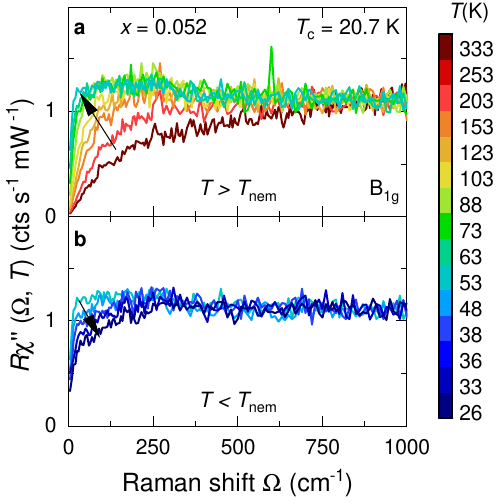}
  \caption{Inelastic light (Raman) scattering data for $x=0.052$ of \bfca in \Bog symmetry. The temperature dependence is limited to frequencies $\Omega < 750\,\mathrm{cm}^{-1}$. \textbf{a} Spectra taken for temperatures $333\,\mathrm{K}>T>T_\mathrm{nem} ~\sim 50\,\mathrm{K}$. At small frequencies, the spectral weight increases sharply towards the nematic transition. \textbf{b} Upon entering the nematic phase, the spectral weight decreases down to the lowest measured temperature at $26\,\mathrm{K}>\Tce$.}
  \label{fig:suppl_data0}
\end{figure}

\begin{figure}[ht]
  \centering
  \includegraphics[width=85mm]{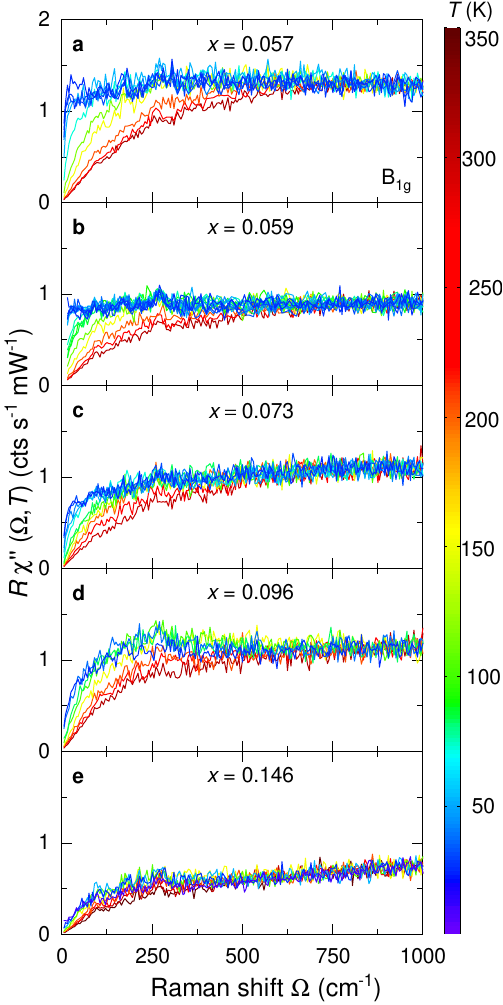}
  \caption{Inelastic light scattering (Raman) data  in \Bog symmetry for doping concentrations as indicated. For all doping concentrations, the small frequency contribution increases with decreasing temperature, but varies qualitatively as the doping $x$ is tuned through and beyond $\xce$ to the overdoped side of the phase diagram.}
  \label{fig:suppl_data1}
\end{figure}

Fig. 2 of the main text shows the correlation function $\beta \tilde{\Lambda} (\beta/2)$ resulting from the transformation of $R\chi''(\Omega, T)$ according to Eq.~(2). Fig.~\ref{fig:suppl_data0} and Fig.~\ref{fig:suppl_data1} compile the spectra which entered this analysis. 

For reasons of clarity, we plot the results for 5.2 \% Co content above \Tnem in panel \textbf{a} and below \Tnem in panel \textbf{b} of Fig.~\ref{fig:suppl_data0}. Here, the pile-up of spectral weight from nematic fluctuations \cite{Kretzschmar:2016} and subsequent collapse for $T<\Tneme$ yield the characteristic cusp \cite{Lederer:2020} of Fig. 2 \textbf{a} in the main text.

Fig.~\ref{fig:suppl_data1} shows the remaining results for $x>0.052$. Here, too, the spectral weight at small frequencies increases as temperature is decreased for all doping concentrations. Yet, this increase is weaker for $x>\xce$ than for $x<\xce$ for which the presence of nematic fluctuations yields a significant contribution to the response. 

\newpage

\section{Scaling}

\noindent In the main text, we have identified quantum critical scaling in the vicinity of optimal doping using the imaginary time correlation function $\tilde\Lambda(\beta/2)$.  More precisely, we have analyzed the temperature dependence of $\beta\tilde\Lambda(\beta/2)$  to extract, for each doping value, a crossover temperature, $T^*$, below which this function begins to deviate from the constant  behavior indicative of a marginal Fermi liquid. Another, more traditional approach is to analyze the real frequency data directly using energy and intensity scaling. In this section we describe this method and its theoretical background, and apply it to the experimental data presented in Fig.~\ref{fig:suppl_data0} and Fig.~\ref{fig:suppl_data1}. The scaling method yields results consistent with the imaginary time analysis, but involves a number of arbitrary methodological choices that hamper the interpretation of the results. 

\subsection{SCALING OF RESPONSE FUNCTIONS}
\vspace{0.5cm}
\noindent Let us consider a system with a quantum tuning parameter $x$ (such as doping) that is near a quantum critical point at $x=x_c$, and let us assume that the system obeys scaling. At each value of $x$, there is a characteristic energy scale $T^*$, which will vanish as $x$ approaches $x_c$. Approaching from the ordered side of the phase diagram, $T^*$ will be (proportional to) the critical temperature, while on the disordered side it will represent a crossover. In both cases, $T^*$ should vanish with critical exponent $\nu z$, where $\nu$ is the correlation length critical exponent and $z$ is the dynamical critical exponent, but this is not relevant to the current discussion. From here on, we will assume that $T,\ T^*$, and  the frequencies ($\Omega$) at which we probe the system are much less than ``UV" scales such as the bandwidth and Fermi energy, so that we are in the low energy regime where quantum criticality can be observed. 

\vspace{0.5cm}
\noindent Within this regime there are two asymptotic limits of temperature. When $T\ll T^*$, the relevant quantum fluctuations (for instance phonons near a structural transition, or paramagnons near a magnetic transition) have energies much greater than the temperature and are thus frozen out. Therefore, these quantum fluctuations produce only quantitative changes to the properties of the system, in comparison to its behavior far from the QCP. As such, this part of the phase diagram is sometimes called the \emph{renormalized classical} regime. By contrast, when $T\gg T^*$, the quantum fluctuations are effectively gapless and therefore produce their most spectacular consequences, such as divergences of thermodynamic quantities with cooling. This is the \emph{quantum critical} regime, typically having a ``fan" shape, as shown in Fig. 3 of the main text.  

\vspace{0.5cm}
\noindent We will discuss these regimes separately below, but first describe the highly constrained form of response functions that we obtain throughout the full range of (small) $T$ and $T^*$ under the assumption of quantum critical scaling. Let $\chi(\Omega)$ be a response function that diverges with characteristic exponent $y$ at the  QCP (if $\chi$ is the order parameter susceptibility, this exponent is $\gamma/\nu z$, where $\gamma$ is the susceptibility critical  exponent), and write $\chi=\chi_0+\chi_s$, where $\chi_0$ is the analytic (smooth) part of the response and $\chi_s$ is the singular part. If the quantum critical point obeys scaling, then $\chi_s$ has the form

\begin{align}
    \chi_s (\Omega;x,T)=|\Omega|^{-y}\cdot {\cal F}\left(\frac{\Omega}{T},\frac{\Omega}{T^*}\right),
    \label{eq:scaling}
\end{align}
where we have suppressed the $x$ dependence of $T^*$ for compactness. In this expression, ${\cal F}$ is a \emph{scaling function}, which is identical for transitions within a given universality class. Except for the characteristic  divergence as $\Omega^{-y}$, the variable $\Omega$ and the parameters $T$ and $ T^*$ appear only as the dimensionless, scaled combinations shown in Eq.~\ref{eq:scaling}. We could write a variety of equivalent expressions involving different scaled variables (for instance, $T/T^*$ instead of $\Omega/T^*$). An example would be $\chi_s=T^{-y}{\cal G}(\frac{\Omega}{T},\frac{T}{T^*})$, with $\cal G$ a different universal scaling function. We have chosen the form above for convenience. The limits of $\cal{F}$  when its arguments go to zero and infinity correspond to the various asymptotic regimes of the  phase diagram and of the dynamical response.  
\subsection{ASYMPTOTIC REGIMES}
We can approach the QCP from the side, with $T=0$ and $T^*>0$, or from the side, with $T>0$ and $T^*=0$. The former case corresponds to the so-called renormalized classical regime, and the latter to the quantum critical regime. In the renormalized classical regime, the first argument of $\cal F$ is infinite, and the response obeys $\Omega/T^* $ scaling:

\begin{align}
\chi_s(\Omega; T =0)=&|\Omega|^{-y}\cdot f_{h}\left(\frac{\Omega}{T^*}\right),\text{ where}\\
f_{h}(u)\equiv &\lim_{v\to\infty}{\cal F}(v,u)
\end{align}
 
Where the subscript ``h" is for ``horizontal." More precisely:

\begin{align}
\chi_s(\Omega; T,T^*)\cdot |\Omega|^{y}\approx f_{h}\left(\frac{\Omega}{T^*}\right),\quad\text{ for } \Omega,T^*\gg T
\end{align}

So, for a given value of $T^*$, the spectra at $T\ll T^*$ depend only on $\Omega$.

\vspace{0.5cm}
\noindent
In the quantum critical regime, the second argument of $\cal F$ is infinite, so the response obeys $\Omega/T $ scaling:

\begin{align}
\chi_s(\Omega; T ^*=0)=&|\Omega|^{-y}\cdot f_{v}\left(\frac{\Omega}{T}\right),\text{ where}\\
f_{v}(u)\equiv &\lim_{v\to\infty}{\cal F}(u,v),
\label{eq:omegaOverT}
\end{align} 

and the subscript ``v" is for  ``vertical." More precisely:

\begin{align}
\chi_s(\Omega; T,T^*)\cdot |\Omega|^{y}\approx f_{v}\left(\frac{\Omega}{T}\right),\quad\text{ for } \Omega,T\gg T^*
\label{scaling}
\end{align} 

So, for a given value of $T^*$, spectra for $T\gg T^*$ depend only on the scale-invariant combination $\Omega/T$.

\subsection{DATA COLLAPSE}

\begin{figure*}[h!]
  \centering
  \includegraphics[width=170mm]{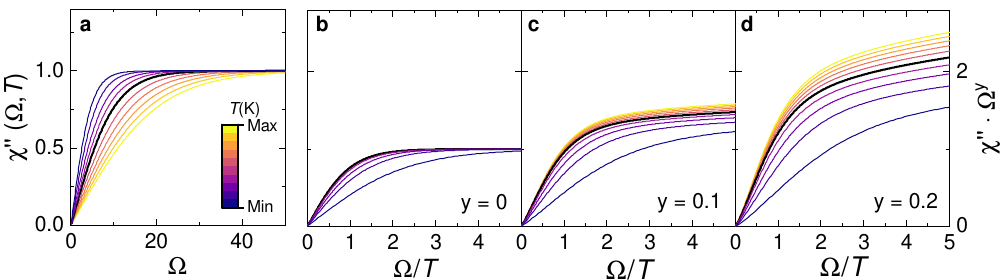}
  \caption{Data collapse for synthetic data with a temperature evolution as indicated. The black curves correspond to $T^*$. \textbf{a}~Down to $T^*$, the response $\chi^{\prime\prime}(\Omega, T)$ grows with $\Omega/T$. Not immediately apparent is the deviation from this scaling for $T<T^*$. \textbf{b}-\textbf{d}~For this idealized data set, data collapse of $\chi^{\prime\prime} \cdot \Omega^y$ plotted against $\Omega/T$ is observed for the high temperature data at $T>T^*$ and $y=0$.}
  \label{fig:suppl_datasynth}
\end{figure*}

\noindent The expressions above suggest a method to estimate the scaling exponent $y$ as well as the crossover temperature $T^*$. By Eq. \ref{scaling} above, $\chi_s\cdot|\Omega|^y$ is a function only of $\Omega/T$ for the high temperature data. Therefore, to the extent that the total response $\chi=\chi_0+\chi_s$ is dominated by the singular part $\chi_s$, the spectra for $T\gg T^*$ and $T\ll T^*$ can be distinguished as follows. First, plot $\chi\cdot \Omega^y$ versus $\Omega/T$ for several values of $y$, and choose the value of $y$ that yields the best collapse of the high temperature spectra. Specializing to this optimal value of $y$, the spectrum measured at the crossover temperature $T^*$ will form a separatrix between the collapsing and non-collapsing spectra. This procedure is shown in Fig.~\ref{fig:suppl_datasynth} for an idealized data set in which $\chi^{\prime\prime}$ grows as $\Omega/T$ for $T>T^*$ and as $\alpha(T)\cdot\Omega/T$ for $T<T^*$. 

\begin{figure*}[h!]
  \centering
  \includegraphics[width=170mm]{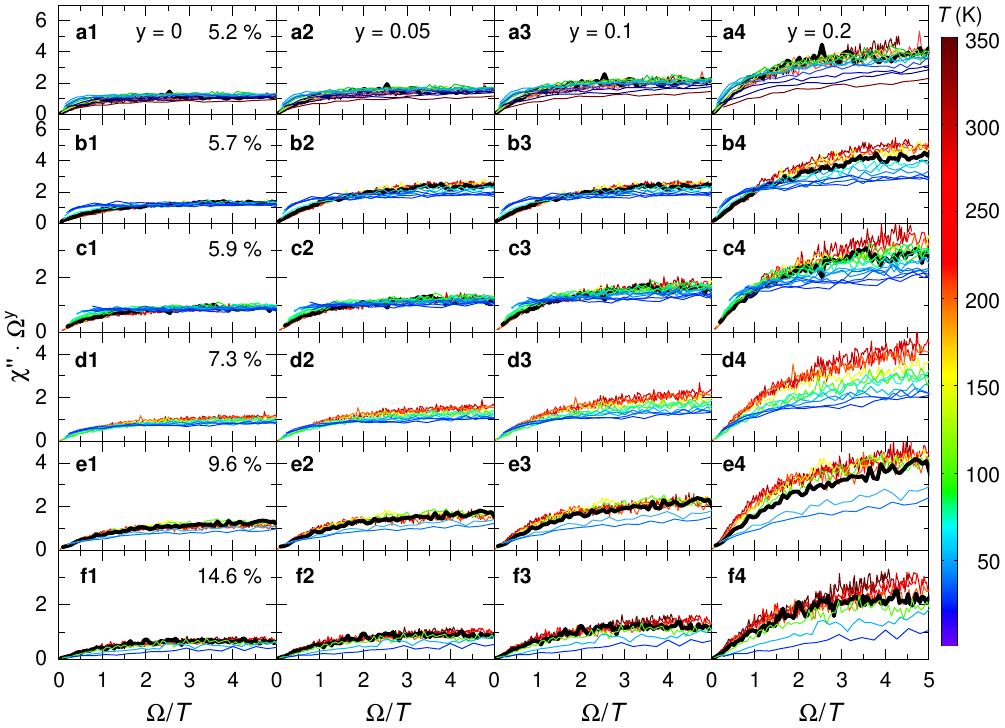}
  \caption{Data collapse method illustrated in the text for varying values of the exponent $y$.}
  \label{fig:suppl_datacollapse}
\end{figure*}

Unfortunately, choosing the best value of $y$ and identifying this separatrix for experimental data must involve some arbitrarily chosen qualitative and/or quantitative criteria, which makes the data collapse procedure somewhat ad hoc. This is in addition to the challenges posed by the regular part $\chi_0$.

\vspace{0.5cm}
\noindent We illustrate the data collapse method using experimental data in Fig.~\ref{fig:suppl_datacollapse}. Upon attempting the collapse for various doping concentrations and various values of $y$, we find that only small (or $0$) values of $y$ show reasonable collapse of the high temperature spectra. This is consistent with the nearly constant behavior of $\beta\tilde\Lambda(\beta/s)$ at high temperatures. For each doping $x$, we highlight in black the spectrum at the value of $T^*(x)$ determined from the imaginary time analysis of the main text. With the exception of $x=7.3\%$, where $T^*$  is below the base temperature of the experiment, the black curves do indeed roughly separate the collapsing and non-collapsing spectra. Therefore, the two methods are consistent, but the imaginary time method provides a much more straightforward way of identifying $T^*$.

\subsection{ASYMPTOTICS OF \BLT}

Here we prove two claims mentioned in the text: The first is that \BLT has the same singular temperature dependence as the static susceptibility upon on cooling towards a quantum critical point exhibiting $\Omega/T$ scaling. To first look at the singularity of $\chi'$ as $T\to 0$ with $T^*=0$, write its singular part as in Eq. \ref{eq:omegaOverT}:

\begin{align}
    \chi_s(\Omega)=&|\Omega|^{-y}\cdot f_{v}\left(\frac{\Omega}{T}\right),
\end{align}

At nonzero temperature, the static susceptibility must be finite, so the divergence of the factor $|\Omega|^{-y}$ at low frequencies must be compensated by the asymptotic behavior of $f_v$. In particular, cancelling the divergence requires $f_v(u)\sim u^y$ as $u\to 0$, so that the static susceptibility diverges with temperature as

\begin{align}
    \chi_s(\Omega=0)\sim & |\Omega|^{-y}\cdot \left(\frac{\Omega}{T}\right)^y\\
    =T^{-y}
\end{align}

Now let us substitute Eq. \ref{eq:omegaOverT} into the expression for \BLT:

\begin{align}
\beta\tilde{\Lambda}_s(\beta/2)=&\int\frac{\mathrm{d}\Omega}{2\pi T} \frac{\chi''(\Omega)}{\sinh[\beta\Omega/2]}\\
=& \int\frac{\mathrm{d}\Omega}{2\pi T} \frac{|\Omega|^{-y}\cdot \mathrm{Im}[f_{v}(\Omega/T)]}{\sinh[\beta\Omega/2]},
\end{align}

where $\tilde\Lambda_s$ is the singular part of $\tilde \Lambda$. Now we rewrite the integral in terms of the scaled variable $x=\Omega/T$, so that $\Omega= x T$:

\begin{align*}
\beta\tilde{\Lambda}_s(\beta/2)=& \int\frac{T\mathrm{d}x}{2\pi T} \frac{|T x|^{-y}\cdot \mathrm{Im}[f_{v}(x)]}{\sinh[x/2]}\\
&\\
=&T^{-y}\cdot\int\frac{\mathrm{d}x}{2\pi} \frac{|x|^{-y}\cdot f_{v}(x)}{\sinh[x/2]},
\end{align*}

The integral is now independent of $T$ and is some order 1 number dependent on the scaling function $f_v$. Therefore, as promised, $\chi'(\Omega=0)$ and \BLT have the same $T^{-y}$ divergence upon cooling towards the QCP.

The second statement in the text is that \BLT generically vanishes as $T\to 0$. By generically we mean that the response function is analytic in frequency, i.e. non-critical. In this case, we can expand $\chi''$ as a convergent Taylor series in $\Omega$:
\begin{equation}
    \chi''(\Omega)=\sum_{n=0}^\infty a_n\Omega^{2n+1},
\end{equation}
where $a_n$ are real coefficients, and only odd powers of $\Omega$ are included because $\chi''$ is odd in frequency. Substitute this into the expression for \BLT:
\begin{align*}
\beta\tilde{\Lambda}_s(\beta/2)=&\int\frac{\mathrm{d}\Omega}{2\pi T\sinh[\beta\Omega/2]} \sum_n a_n \Omega^{2n+1}
\end{align*}
Now once again substitue $\Omega=xT$

\begin{align*}
\beta\tilde{\Lambda}_s(\beta/2)=&\int\frac{T\mathrm{d}x}{2\pi T\sinh[x/2]} \sum_n a_n T^{2n+1}x^{2n+1}\\
=&\sum_n \alpha_n T^{2n+1} ,
\end{align*}

where

\begin{align*}
\alpha_n=&a_n\int\frac{\mathrm{d}x}{2\pi\sinh[x/2]} x^{2n+1}
\end{align*}
And each of the $\alpha_n$ are convergent because of the exponential cutoff. Accordingly, \BLT can be written as a series in odd powers of $T$, and therefore vanishes as $T\to 0$. The only way this can fail to occur is if the response diverges sufficiently rapidly at low frequencies.

\section{Sample characterization}
\begin{figure}[h!]
  \centering
  \includegraphics[width=85mm]{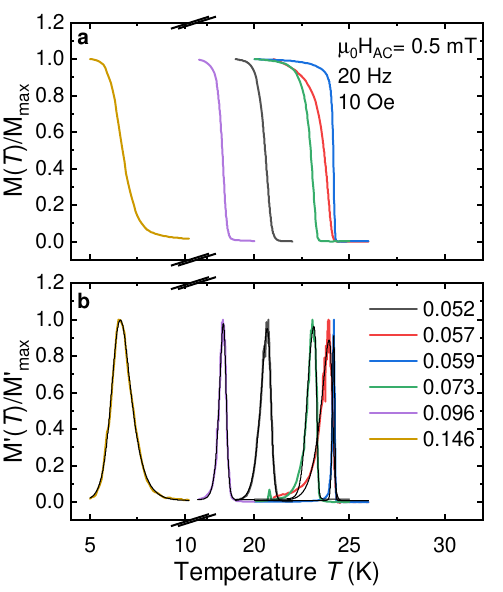}
  \caption{\textbf{a} Normalized magnetization as a function of temperature for doping concentrations as indicated. The increase from the high temperature side signifies the onset of the superconducting transition which saturates in the superconducting state. \textbf{b} First derivative of the magnetization $\partial \mathrm{M}(T)/\partial T$ normalized to the peak maxima alongside the fits in black described in the text. }
  \label{fig:suppl_squid}
\end{figure}

The samples were cleaved \textit{ex-situ} prior to mounting the sample holder into the cryostat. Characterization with a superconducting quantum interference device (SQUID) was performed after completion of the Raman measurements. The results are presented in Fig.~\ref{fig:suppl_squid} \textbf{a}. Assuming an S-shaped function for the magnetization curves of the form

\begin{equation}
M[T] \sim  \frac{1}{1+\mathrm{exp}[-(T-\Tce)/\Gamma^{(1)}]} \cdot  \frac{1}{1+\mathrm{exp}[-(T-\Tce)/\Gamma^{(2)}]}
\end{equation}

we derive the critical temperatures \Tc of the samples from the peak positions of $\partial \mathrm{M}(T)/\partial T$ from a standard asymmetric double sigmoidal function as a fit model (black curves in Fig.~\ref{fig:suppl_squid} \textbf{b}). The results are compiled in Table~\ref{tab:samples}.

\begin{table}[h!]
\centering
\caption{Sample characterization. Doping concentrations $x$ and their variation within each sample $\Delta x$ derived from the width $\pm \Delta \Tce$ of $\partial \mathrm{M}(T)/\partial T$. \Tc corresponds to the peak position.}
\label{tab:samples}
\begin{tabular}{||c| c| c| c||} 
 \hline
Doping $x$ (\%) &  Variance $\Delta x$ (\%) & $T_\mathrm{c}$ (K) & $\pm\Delta T_\mathrm{c}$ (K) \\ [0.5ex] 
 \hline\hline
5.2 & 0.15 & 20.7 & 0.2 \\ 
 \hline
5.7 & 0.05 & 23.9 & 0.2 \\
 \hline
5.9 & 0.05 & 24.2 & 0.05 \\
 \hline
7.3 & 0.13 & 23.0 & 0.13 \\
 \hline
9.6 & 0.05 & 18.4 & 0.10\\ [1ex] 
 \hline
14.6 & 0.16 & 6.4 & 0.25\\ [1ex] 
 \hline
\end{tabular}
\end{table}

\clearpage
\newpage

\bibliographystyle{unsrtnat}
\bibliography{Jost_QF_bib}

\end{document}